\documentclass[conference]{IEEEtran}
\IEEEoverridecommandlockouts
\usepackage{cite}
\usepackage{amsmath,amssymb,amsfonts}
\usepackage{mathtools}
\usepackage{algorithmic}
\usepackage{array}
\usepackage[pdftex]{graphicx}
\graphicspath{ {./images/} }
\usepackage{textcomp}
\usepackage{xcolor}
\def\BibTeX{{\rm B\kern-.05em{\sc i\kern-.025em b}\kern-.08em
    T\kern-.1667em\lower.7ex\hbox{E}\kern-.125emX}}
\usepackage{nomencl}
\usepackage{etoolbox}
\renewcommand\nomgroup[1]{%
  \item[\bfseries
  \ifstrequal{#1}{B}{Indexes and Sets}{%
  \ifstrequal{#1}{C}{Functions}{%
  \ifstrequal{#1}{D}{Parameters}{%
  \ifstrequal{#1}{A}{Abbreviations}{%
  \ifstrequal{#1}{E}{Time-series}{}}}}}%
]}
\makenomenclature

\usepackage[]{fancyhdr} %
\newcommand{\changefont}{\fontsize{9}{9}\selectfont}
\begin{document}

\title{Estimating the Future Need of Balancing Power Based on Long-Term Power System\\
Market Simulations}

\author{\IEEEauthorblockN{Henrik Nordström, Lennart Söder}
\IEEEauthorblockA{Electric Power and Energy Systems \\
\textit{KTH Royal Institute of Technology} \\
Stockholm, Sweden \\
hnordstr@kth.se, lsod@kth.se}
\and
\IEEEauthorblockN{Robert Eriksson}
\IEEEauthorblockA{System Development\\
\textit{Svenska Kraftnät}\\
Sundbyberg, Sweden \\
robert.eriksson@svk.se}
}

\maketitle
\thispagestyle{fancy}
\pagestyle{fancy}

\begin{abstract}
With increasing penetration of variable renewable energy sources (vRES) and a higher rate of electrification in the society, there will be future challenges in maintaining a continuous balance between electricity supply and demand. To investigate the possibility of different technologies to provide balancing services, to dimension future balancing services and design technical requirements for future balancing services, the future \textit{need of balancing power} must first be known. Future power systems are often analysed by performing simulations capturing the energy produced, consumed or transmitted for different power system components with resolution of one trading period (TP), here called \textit{TP energy simulations}. However, these simulations do not fully capture the continuous intra-TP power balance that must be kept at every time instance. In this paper, we propose a model to perform \textit{high-resolution intra-TP simulations} to estimate the \textit{need of balancing power} based on \textit{TP energy simulations}. The model is applied to the Swedish system operator (SO) Svenska kraftnät's \textit{TP energy simulations} of a scenario with high vRES penetration and a highly electrified society in year 2045. The results show that a considerable \textit{need of balancing power} will occur frequently, where faster ramping of components tend to increase the \textit{need of balancing power} while assuming that the transmission reliability margin (TRM) is used to \textit{net imbalances} clearly decreases the \textit{need of balancing power}. 
\end{abstract}


\nomenclature[A,1]{AC}{Alternating current}
\nomenclature[A,2]{ACE}{Area control error}
\nomenclature[A,3]{HVDC}{High voltage direct current}
\nomenclature[A,4]{NTC}{Net transfer capacity}
\nomenclature[A,5]{SO}{System operator}
\nomenclature[A,6]{TP}{Trading period}
\nomenclature[A,7]{TRM}{Transmission reliability margin}
\nomenclature[A,8]{vRES}{Variable renewable energy sources}

\nomenclature[B,1]{$A$}{Set of AC interconnections, indexed $n1n2$}
\nomenclature[B,2]{$B$}{Set of HVDC interconnections, indexed $n1n2$}
\nomenclature[B,3]{$i$}{Integer, $i \in (1, \dots ,I)$}
\nomenclature[B,4]{$n$}{Trading areas, $n \in (1,\dots,N)$}
\nomenclature[B,5]{$t$}{TP resolution, $t \in (1,\dots,T)$}
\nomenclature[B,6]{$\hat{t}$}{Needed higher resolution, $\hat{t} \in (1,\dots,\hat{T})$}

\nomenclature[C,1]{$f(*)$}{A method to create high-resolution data, could be $HR_{C}(*)$ or $HR_{V}(*)$}
\nomenclature[C,2]{$HR_{C}(*)$}{A method to create high-resolution data for controllable components}
\nomenclature[C,3]{$HR_{V}(*)$}{A method to create high-resolution data for varying components}

\nomenclature[D,01]{$\alpha$}{Cost of deviation of transmission}
\nomenclature[D,02]{$C_{t \rightarrow t+1}$}{Parameter determining ramping period between TP $t$ and $t+1$ [min]}
\nomenclature[D,03]{$e^{i}$}{The total error for a component in iteration $i$}
\nomenclature[D,04]{$e_{min}$}{An accepted total error}
\nomenclature[D,05]{$h_{t}^{i}$}{The difference in TP energy between high-resolution data and \textit{TP energy simulation} data during TP $t$ in iteration $i$ [MWh]}
\nomenclature[D,06]{$r$}{Ramp rate [MW/min] or [\%/min]}
\nomenclature[D,07]{$\overline{Z}^{ac}_{n1n2}$}{Maximum positive transmission capacity of AC-interconnection $n1n2$ [MW]}
\nomenclature[D,08]{$\underline{Z}^{ac}_{n1n2}$}{Maximum negative transmission capacity of AC-interconnection $n1n2$ [MW]}

\nomenclature[E,01]{$a_{t}^{i}$}{A variable time-series for a component in iteration $i$ of $t$-resolution [MWh/TP]}
\nomenclature[E,02]{$d_{t,n}$}{Demand in area $n$ of $t$-resolution [MWh/TP]}
\nomenclature[E,03]{$\hat{d}_{\hat{t},n}$}{Demand in area $n$ of $\hat{t}$-resolution [MW]}
\nomenclature[E,04]{$g^{fl}_{t,n}$}{Flexible production in area $n$ of $t$-resolution [MWh/TP]}
\nomenclature[E,05]{$\hat{g}^{fl}_{\hat{t},n}$}{Flexible production in area $n$ of $\hat{t}$-resolution [MW]}
\nomenclature[E,06]{$g^{hy}_{t,n}$}{Hydro production in area $n$ of $t$-resolution [MWh/TP]}
\nomenclature[E,07]{$\hat{g}^{hy}_{\hat{t},n}$}{Hydro production in area $n$ of $\hat{t}$-resolution [MW]}
\nomenclature[E,08]{$g^{nu}_{t,n}$}{Nuclear production in area $n$ of $t$-resolution [MWh/TP]}
\nomenclature[E,09]{$\hat{g}^{nu}_{\hat{t},n}$}{Nuclear production in area $n$ of $\hat{t}$-resolution [MW]}
\nomenclature[E,10]{$g^{res}_{t,n}$}{vRES production in area $n$ of $t$-resolution [MWh/TP]}
\nomenclature[E,11]{$\hat{g}^{res}_{\hat{t},n}$}{vRES production in area $n$ of $\hat{t}$-resolution [MW]}
\nomenclature[E,12]{$g^{th}_{t,n}$}{Thermal production in area $n$ of $t$-resolution [MWh/TP]}
\nomenclature[E,13]{$\hat{g}^{th}_{\hat{t},n}$}{Thermal production in area $n$ of $\hat{t}$-resolution [MW]}
\nomenclature[E,14]{$w_{t}$}{A data-series from a \textit{TP energy simulation} of $t$-resolution [MWh/TP]}
\nomenclature[E,15]{$\hat{w}_{\hat{t}}^{i}$}{A high-resolution data series of a component in iteration $i$ of $\hat{t}$-resolution [MW]}
\nomenclature[E,16]{$\hat{w}^{act}_{\hat{t}}$}{Actual power of $\hat{t}$-resolution [MW]}
\nomenclature[E,17]{$\hat{w}_{\hat{t},n}^{bal}$}{Need of balancing power in area $n$ of $\hat{t}$-resolution [MW]}
\nomenclature[E,18]{$\hat{w}^{bas}_{\hat{t}}$}{Basic power of $\hat{t}$-resolution [MW]}
\nomenclature[E,19]{$\hat{w}^{imb}_{\hat{t}}$}{Power imbalance of $\hat{t}$-resolution [MW]}
\nomenclature[E,20]{$w^{imb}_{t}$}{TP energy imbalance of $t$-resolution [MWh/TP]}
\nomenclature[E,21]{$w^{TP}_{t}$}{Energy of one component from TP energy simulations of $t$-resolution [MWh/TP]}
\nomenclature[E,22]{$z^{ac}_{t,n1n2}$}{AC transmission from area $n1$ to $n2$ of $t$-resolution [MWh/TP]}
\nomenclature[E,23]{$\hat{z}^{ac}_{\hat{t},n1n2}$}{AC transmission from area $n1$ to $n2$ of $\hat{t}$-resolution [MW]}
\nomenclature[E,24]{$z^{dc}_{t,n1n2}$}{HVDC transmission from area $n1$ to $n2$ of $t$-resolution [MWh/TP]}
\nomenclature[E,25]{$\hat{z}^{dc}_{\hat{t},n1n2}$}{HVDC transmission from area $n1$ to $n2$ of $\hat{t}$-resolution [MW]}
\printnomenclature

\section{Introduction}
\begin{figure}[tp]
\centerline{\includegraphics[width=0.85\linewidth]{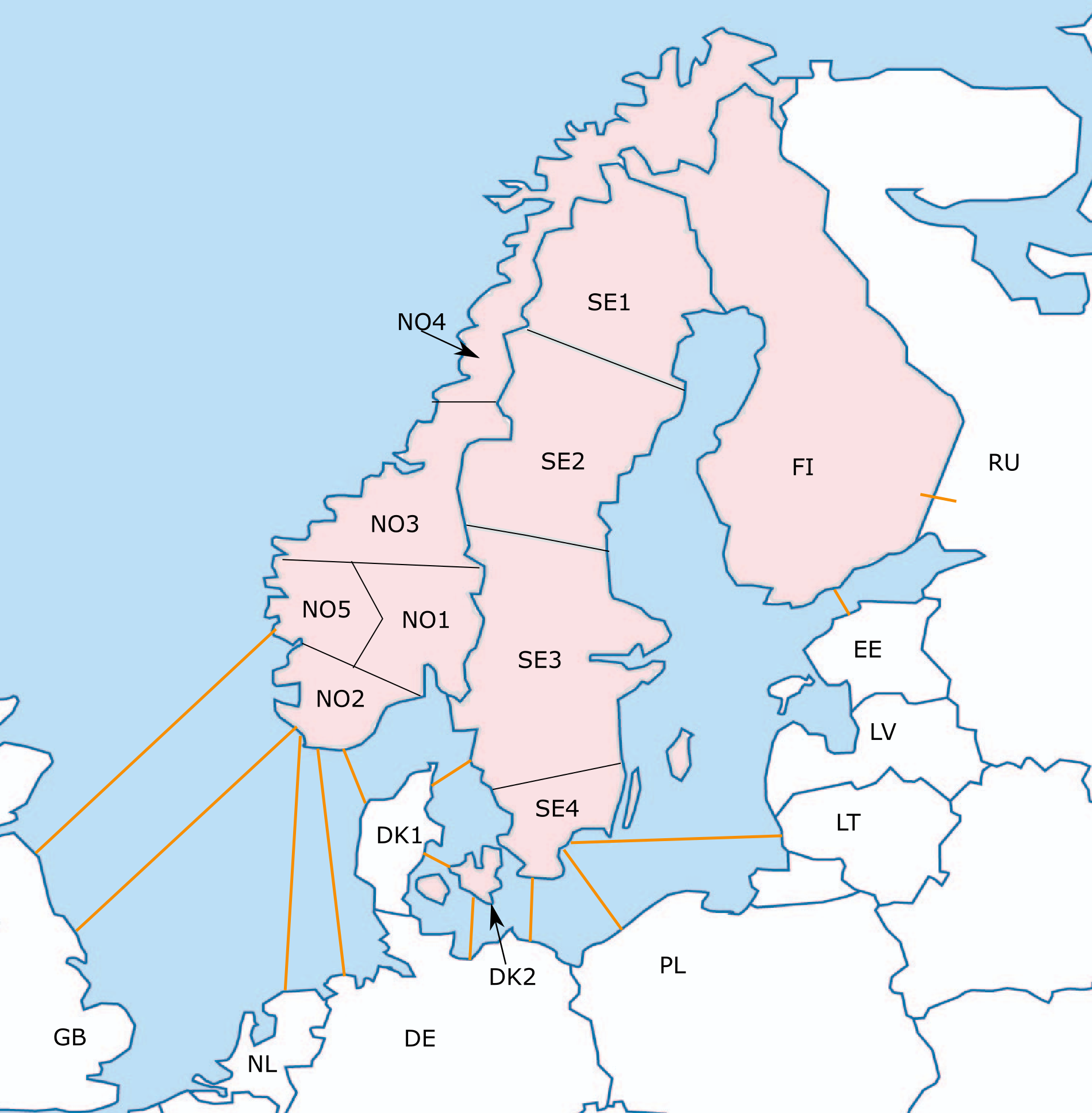}}
\caption{The Nordic synchronous area (pink) and all HVDC interconnections (orange lines) to external areas as planned in year 2045.}
\label{fig:nordic}
\end{figure}

Alarming reports about the potential consequences of global warming stress the need of action to rapidly mitigate global greenhouse gas emissions \cite{b1}. A key pillar to keep the global warming below 1.5$^\circ$C in accordance with the Paris agreement is to replace energy from burning fossil fuels with electricity produced by renewable energy sources \cite{b2}. However, this will impose new challenges to power systems as 1) the electricity production patterns will change due to the penetration of variable renewable energy sources (vRES) like wind and solar and 2) the electricity demand patterns will change as sectors like the industrial sector and the transport sector will use substantial amounts of electricity when phasing out fossil fuels. The Nordic power system is expected to go through substantial changes in the strive towards climate neutrality with about a 122\% increase in installed vRES capacity between years 2020 and 2040 leading to a more volatile system where flexible resources such as demand side response, power-to-x, storage and electric vehicles will be important to even out variations in vRES production \cite{b3}. Also, the amount of HVDC interconnections between the Nordic synchronous power system and external power systems are expected to increase. The Nordic synchronous system is already interconnected to three different synchronous systems (the Russian/East European system, the Central European system and United Kingdom's system). A map of the eleven trading areas in the Nordic synchronous power system and the HVDC interconnections to external areas as planned in 2045 is shown in Fig.~\ref{fig:nordic}. A new balancing model for the Nordic countries aiming to harmonize with other European power systems with new ways of procuring and activating balancing services as well as introducing 15 minute trading period (TP) is currently (2022) being implemented with the aim of being fully implemented in year 2024 \cite{b4}.

In \cite{b5}, the adequacy of the Nordic power system is analysed for different development scenarios in year 2035 and 2045 with emphasis on the Swedish power system. To maintain safe future operations of the power system, flexible resources providing balancing services are mentioned as a key if Sweden fulfills its target of 100\% renewable electricity generation in year 2040 while still enabling a society with a high rate of electrification. Hence, there will be an increased \textit{need of balancing power} in a future Nordic power system based on renewable energy sources caused by both variability as well as uncertainty in forecasts. By knowing the \textit{need of balancing power}, further analysis of how balancing services should be dimensioned, how technical requirements of balancing services should be designed and which technologies could provide cost-efficient balancing services can be performed. A structural approach to learn about the \textit{need of balancing power} in a power system with an energy-only market where mainly data of energy per TP is available or studied (like the Nordic power system) is to 1) study the intra-TP power system dispatch to understand challenges caused by variability and 2) study the forecast errors of production and demand to understand the challenges caused by uncertainty. This paper deals with the first task and we present a model aiming to provide clarity regarding the intra-TP power system dispatch in studies of future power systems.
\newline

To analyse future needs and challenges of power systems, market simulations of different development scenarios for power systems are often run by system operators (SOs) or other institutions, e.g. \cite{b5}, \cite{b6}, \cite{b7} and \cite{b8}. In Europe, best practice for these simulations is to use area models simulating the aggregated electricity production and demand in larger geographical areas (trading areas) as well as the transmission between these trading areas. In North America, best practice is instead to use nodal models which will give more detailed results but also require more input data. Both area- and nodal-based simulations are often performed by running unit commitment and economic dispatch with time resolution of one TP to find out how the energy demand in each area or node will be met in the most cost-efficient way for the entire power system. Thus we call these type of simulations \textit{TP energy simulations}. The power system adequacy is captured by studying the power systems' capability to meet the energy demand in each area or node every TP. By constraining the simulations to keep certain levels of production reserves and margins on transmission lines, fluctuations within each TP are assumed to be handled by the power system.
\newline

In \cite{b9} the authors investigated the impact of modelling with finer time resolution by running unit commitment and economic dispatch of the Irish power system for year 2020 with time resolutions from 60 minutes down to 5 minutes. The authors found high-resolution simulations beneficial when the flexibility of the system in terms of ramping and flexible resources is of interest, which is the case for power systems with high penetration of vRES as there will be larger challenges to maintain a continuous balance between supply and demand. If performing simulations with low time resolution like 60 minutes, it is hard to capture the actual ramping of components and it is most often only modelled as a maximum allowed change in energy between time steps. However, there may be computational challenges in running unit commitment and economic dispatch with 5 minutes resolution of a detailed model of larger power systems including multiple countries and synchronous areas. 
\newline

In line with the findings in \cite{b9}, North American vRES integration studies like \cite{b7} and \cite{b8} perform sub-hourly simulations by running economic dispatch with 5 minutes resolution by following a certain procedure. First, unit commitment and economic dispatch is run with 60 minutes resolution based on historical weather forecast data. In this step, the reserves for each hour are procured based on the load and the vRES production forecasted for each hour by requiring generation available in online generators. This step represents the clearing of the day-ahead market. Second, economic dispatch is run with 5 minutes resolution. Commitment schedules from the 60 minutes resolution simulation are used in combination with an updated time series of weather data, interpreted as the "actual" weather. In this step, online generators are re-dispatched and the capacity reserved for reserves is re-optimized. This can be seen as simulating the real-time market. Thus, this approach is able to capture if enough generation is available to meet the need of regulation and contingency reserves when considering that online generators must change their generation due to forecast errors. However, this approach both require knowledge regarding the relation between the need of reserves and the load and vRES production as well as knowledge regarding which generators will provide reserves. As the Nordic power system is going through substantial changes, it is yet unclear how the need of different reserves is determined as well as which resources will provide the needed reserves. By first analysing the \textit{need of balancing power}, clarity regarding these concerns may be provided and high-resolution economic dispatch simulations made possible.
\newline

In \cite{b10}, an approach to model the power system balancing under the uncertainty of vRES production with minutely resolution is presented. First reserves are scheduled by running a security-constrained unit commitment of hourly resolution to clear the day-ahead market. Then the scheduled reserves are managed in real-time by measuring the power system state and imbalances by power flow analysis, activating regulation services in minutely resolution while re-dispatching generators in the real-time market by running a security-constrained economic dispatch with 5 minutes resolution in parallel. If the power system imbalances still exceed a pre-determined level, manual actions will be activated. However, this method requires both vRES forecast data of hourly and minutely resolution to clear the day-ahead and real-time markets as well as actual vRES data to determine the imbalances. Also, load is assumed to have no forecast errors and all the imbalances are caused by vRES production. Thus, the aim is rather to evaluate if methods to schedule and activate reserves are adequate to deal with vRES production forecasts error than to estimate the \textit{need of balancing power} to maintain a continuous balance between supply and demand.
\newline

Methods for short-term estimation of future imbalances based on machine learning algorithms are presented in \cite{b11} and \cite{b12}. These methods are based on using historical imbalance data together with forecasted power system dispatch to estimate the future imbalances. However, data of historical imbalances are not adequate to estimate the long-term future power system imbalances as the imbalance patterns may substantially change if the characteristics of electricity generation and demand changes. In \cite{b13}, a method is presented to estimate the structural imbalances in the Nordic power system based on market simulations of future power systems with emphasis on the impact of HVDC ramp rates and a change to a 15 minutes TP. This method is based on using interpolation methods to create high-resolution data from \textit{TP energy simulation} data. However, this method simplifies by not considering transmission between trading areas in the Nordic synchronous system as well as aggregating multiple trading areas in the same country to one node. Thus, the method does not capture how internal congestions in the Nordic synchronous system affect the imbalances. In \cite{b14} and \cite{b15}, interpolation methods are used to create data with resolution of one second based on historical hourly data for each of the eleven trading areas in the Nordic synchronous power system to enable estimation of the system frequency and evaluation of different balancing strategies. However, transmission between trading areas is not taken into consideration and the method requires historical frequency measurements to properly tune model parameters. 
\newline

Hence, to the best of our knowledge, there are no earlier publications that present a method to perform high-resolution simulations based on \textit{TP energy simulations} to enable estimation of the \textit{need of balancing power} caused by variability that 1) solely relies on the \textit{TP energy simulation} data and a few technical assumptions regarding the studied power system, hence not requiring a lot of different input data 2) creates high-resolution data with realistic patterns in a computationally fast and simple way and 3) considers the \textit{netting of imbalances} by using transmission between nodes. By knowing the future \textit{need of balancing power}, the possibility of using the flexibility of technologies like electric vehicles, hydro power, heat pumps, battery storage and gas turbines to keep the continuous balance in a power system can be evaluated. Also, the \textit{need of balancing power} could be used for dimensioning balancing services and designing technical requirements for balancing services in a future power system. The contribution of this paper is the proposal of a new model to create high-resolution data based on \textit{TP energy simulations} to estimate the \textit{need of balancing power} caused by the variability of electricity production and demand as well as ramping of components and transmission limitations. The aim of the model is to provide clarity regarding how the intra-TP power system dispatch may look in \textit{TP energy simulations}. The model is applied to the scenario "Electrification renewables year 2045" in \cite{b5}, by using simulation results provided by the Swedish SO Svenska kraftnät. The model builds upon the work in \cite{b14} and \cite{b15}, but is further developed by also capturing the high-resolution transmission between trading areas and is customised to be applied for studies of future power systems with \textit{TP energy simulation} data as input. 
\newline

The rest of the paper continues as follows:
In section \ref{sec:definitions}, we define the terminology regarding \textit{imbalances} used in this paper. In section \ref{sec:model}, we describe the setup of the model used to create high-resolution data. In section \ref{sec:case}, we describe the case study to which the model was applied. In section \ref{sec:results}, we present the results of the case study which are then discussed in section \ref{sec:discussion}. In section \ref{sec:conclusions}, we draw some concluding remarks regarding this paper while we present some areas of future work in section \ref{sec:future}.

\section{Definitions}\label{sec:definitions}
As the term \textit{imbalance} is wide and might refer to several different events in a power system context, there is a need to clearly define the terminology used regarding \textit{imbalances} and power system balancing in this paper. Simulations with a resolution of one TP like \cite{b5}, e.g. of hourly resolution, are referred to as \textit{TP energy simulations}. The \textit{TP energy simulations} often have different geographical granularities like "nodes", "zones", "buses", "trading areas", "countries" etc. In this section, we call the geographic fragment in a \textit{TP energy simulation} a node. A data series in a \textit{TP energy simulation} is referred to as a component, i.e. transmission on one interconnection between two nodes, one type of production in a node or the consumption in a node.  When referring to "actual power" or "actual energy" in this section, we mean the outcome of \textit{high-resolution intra-TP simulations}. Although they are in fact just simulated values, we interpret them as actual power measurements when analysing the results of the \textit{high-resolution intra-TP simulations}. 
\newline

The \textit{basic power}, $\hat{w}^{bas}$, is defined as the assumed constant power of one certain component within the TP in \textit{TP energy simulations}. As \textit{TP energy simulations} simulate the energy in certain time intervals rather than the power at certain time instants, it is assumed that each component will provide a constant power within each time interval leading to the simulated energy being supplied, consumed or transmitted. Hence, the power of a component will follow a "step function" with each step occurring at shifts of TP. If every single component delivers the \textit{basic power} within each TP according to a \textit{TP energy simulation}, there will be a continuous balance between supply and demand in every node. In reality, it is of course well-known that most components will not provide a constant power within each TP, instead this is handled by keeping reserves and margins to maintain a continuous supply-demand balance. However in the \textit{TP energy simulations} as well as in the clearing of energy-only markets, the assumption of every component providing constant power is made. Also, deviations from the \textit{basic power} are neither formally restricted nor penalized in the Nordic market as long as the energy provided in a TP equals the amount of energy bidded. By using (\ref{eq:basic}), the \textit{basic power} can be computed for a certain component during one TP, $t$, based on the energy in the \textit{TP energy simulations}, $w^{TP}$. $T$ is the number of data points from the \textit{TP energy simulations} and $\hat{T}$ is the amount of data points from the \textit{high-resolution intra-TP simulations} during the simulated period. 

\begin{equation}\label{eq:basic}
\hat{w}^{bas}_{\hat{t} = 1 + (t-1) \frac{\hat{T}}{T}}, \dots ,\hat{w}^{bas}_{\hat{t} = t \frac{\hat{T}}{T}} = \frac{w^{TP}_{t} \cdot 60}{\hat{T} / T}
\end{equation}

A \textit{power imbalance}, $\hat{w}^{imb}$, is defined as the instantaneous deviation between the actual power, $\hat{w}^{act}$, and the \textit{basic power} for one certain component. In this paper we interpret data of minutely resolution as an "instantaneous" power measurement, although it in fact represents the energy during one minute. Eq.~(\ref{eq:pimb}) shows how the \textit{power imbalance} for one component at time $\hat{t}$ is computed.

\begin{equation}\label{eq:pimb}
\hat{w}^{imb}_{\hat{t}} = \hat{w}^{act}_{\hat{t}} - \hat{w}^{bas}_{\hat{t}}
\end{equation}

\begin{figure}[htbp]
\centerline{\includegraphics[width=\linewidth]{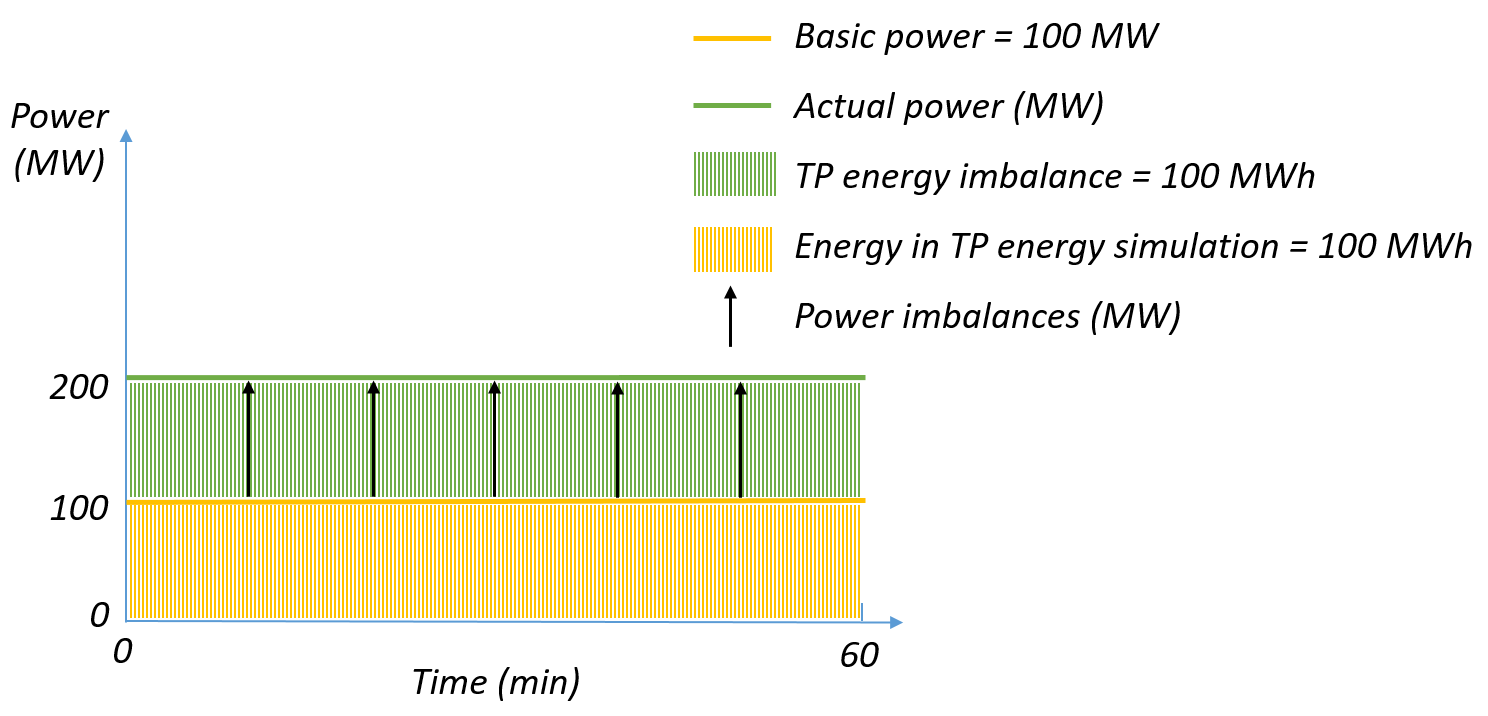}}
\caption{An example of \textit{power imbalances} causing a \textit{TP energy imbalance}.}
\label{fig:TPimb}
\end{figure}

\begin{figure}[htbp]
\centerline{\includegraphics[width=\linewidth]{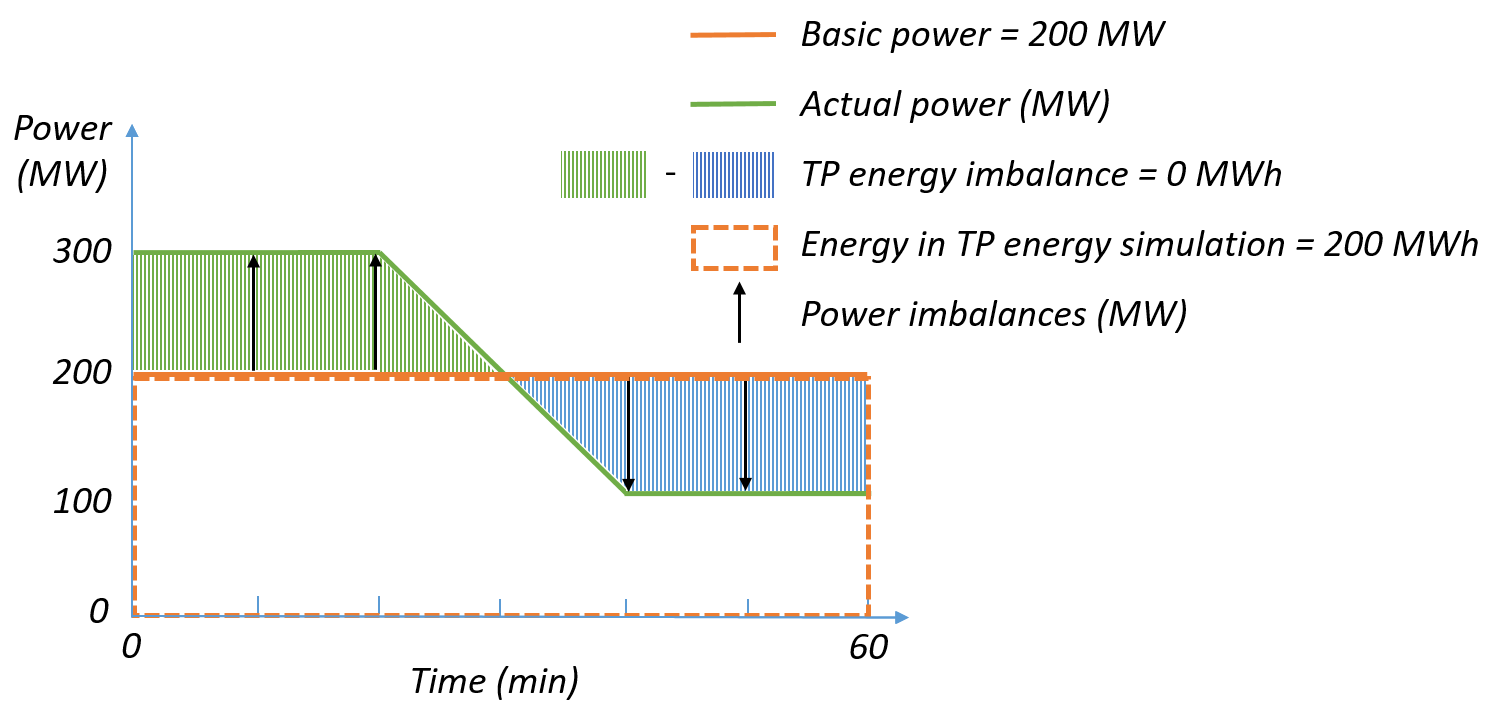}}
\caption{An example of \textit{power imbalances} occurring without causing a \textit{TP energy imbalance}.}
\label{fig:powerimb}
\end{figure}

A \textit{TP energy imbalance}, $w^{imb}_{t}$, is defined as the deviation between actual energy in one TP and the energy in \textit{TP energy simulations} during the same TP for one certain component. \textit{TP energy imbalances} that occur in reality are caused by imperfect forecasts. However, in this paper we assume perfect forecasts in the \textit{TP energy simulations} and hence no \textit{TP energy imbalances} will occur. This assumption is explained by the main purpose of the model being to provide clarity regarding the intra-TP power system behaviour in \textit{TP energy simulations}.  Note that \textit{power imbalances} are the cause of \textit{TP energy imbalances}, but the occurrence of \textit{power imbalances} must not imply that a \textit{TP energy imbalance} will occur. Fig.~\ref{fig:TPimb} shows an example of \textit{power imbalances} causing a \textit{TP energy imbalance} and Fig.~\ref{fig:powerimb} shows how \textit{power imbalances} can occur without causing a \textit{TP energy imbalance}. The \textit{TP energy imbalance} can be computed with (\ref{eq:TPimb}) for one TP, $t$, based on the \textit{power imbalances} of resolution $\hat{t}$ for one certain component.

\begin{equation}\label{eq:TPimb}    
w^{imb}_{t} = \frac{T}{\hat{T}} \sum_{\mathclap{\hat{t}=\frac{\hat{T}}{T}(t-1)+1}}^{\frac{\hat{T}}{T}t} \hat{w}^{imb}_{\hat{t}}
\end{equation}

The \textit{need of balancing power}, $\hat{w}^{bal}$, is defined as the additional power needed to be injected to or absorbed by a node to have an instantaneous supply-demand balance in that certain node. The \textit{need of balancing power} is caused by a lack of coherence between \textit{power imbalances} from production, consumption and transmission in a node. Hence, the \textit{need of balancing power} is equal to the \textit{power imbalances} of consumption and export subtracted with the \textit{power imbalances} of production and import in a node. A positive \textit{need of balancing power} indicates that power must be injected to a node to make up for an instantaneous power deficit and a negative \textit{need of balancing power} indicates that power must be absorbed by a node to make up for an instantaneous power surplus. Comparing the \textit{need of balancing power} to the more commonly used area control error (ACE), the ACE is the deviation for secondary and tertiary control to handle while the \textit{need of balancing power} is the total deviation for primary, secondary and tertiary control to handle \cite{b16}\cite{b17}. Given the data categories in the \textit{TP energy simulations} in \cite{b5}, the \textit{need of balancing power} in node $n$ during time $\hat{t}$ is computed according to (\ref{eq:balpow}). 

\begin{align}\label{eq:balpow}
&\hat{w}_{\hat{t},n}^{bal} = - \hat{g}^{hy}_{\hat{t},n} - \hat{g}^{fl}_{\hat{t},n} - \hat{g}^{th}_{\hat{t},n} - \hat{g}^{nu}_{\hat{t},n} - \hat{g}^{res}_{\hat{t},n} + \hat{d}_{\hat{t},n} \\
&+ \sum_{\mathclap{a \in \{A:n1=n\}}}\hat{z}^{ac}_{\hat{t},a} \text{\hspace{2pt}} - \text{\hspace{2pt}} \sum_{\mathclap{a \in \{A:n2=n\}}}\hat{z}^{ac}_{\hat{t},a} \text{\hspace{2pt}} + \text{\hspace{2pt}} \sum_{\mathclap{b \in \{B:n1=n\}}}\hat{z}^{dc}_{\hat{t},b} \text{\hspace{2pt}}  - \text{\hspace{2pt}} \sum_{\mathclap{b \in \{B:n2=n\}}}\hat{z}^{dc}_{\hat{t},b} \nonumber \\
\nonumber
\end{align}

\begin{figure}[htp]
\centerline{\includegraphics[width=0.9\linewidth]{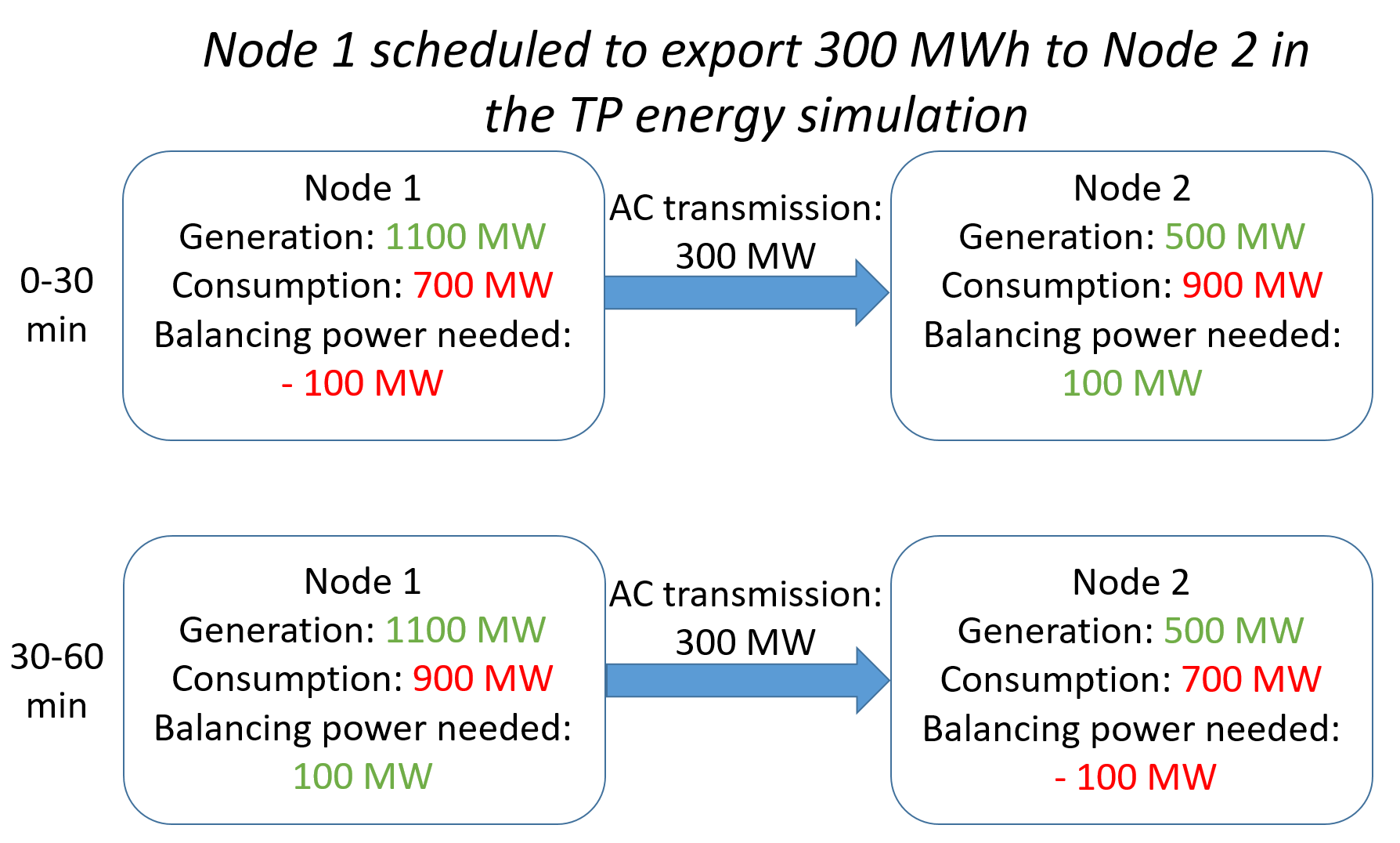}}
\caption{An example of the \textit{need of balancing power} in a simple two node system during one TP when transmission is not used to \textit{net imbalances}.}
\label{fig:prenetting}
\end{figure}

\begin{figure}[htp]
\centerline{\includegraphics[width=0.9\linewidth]{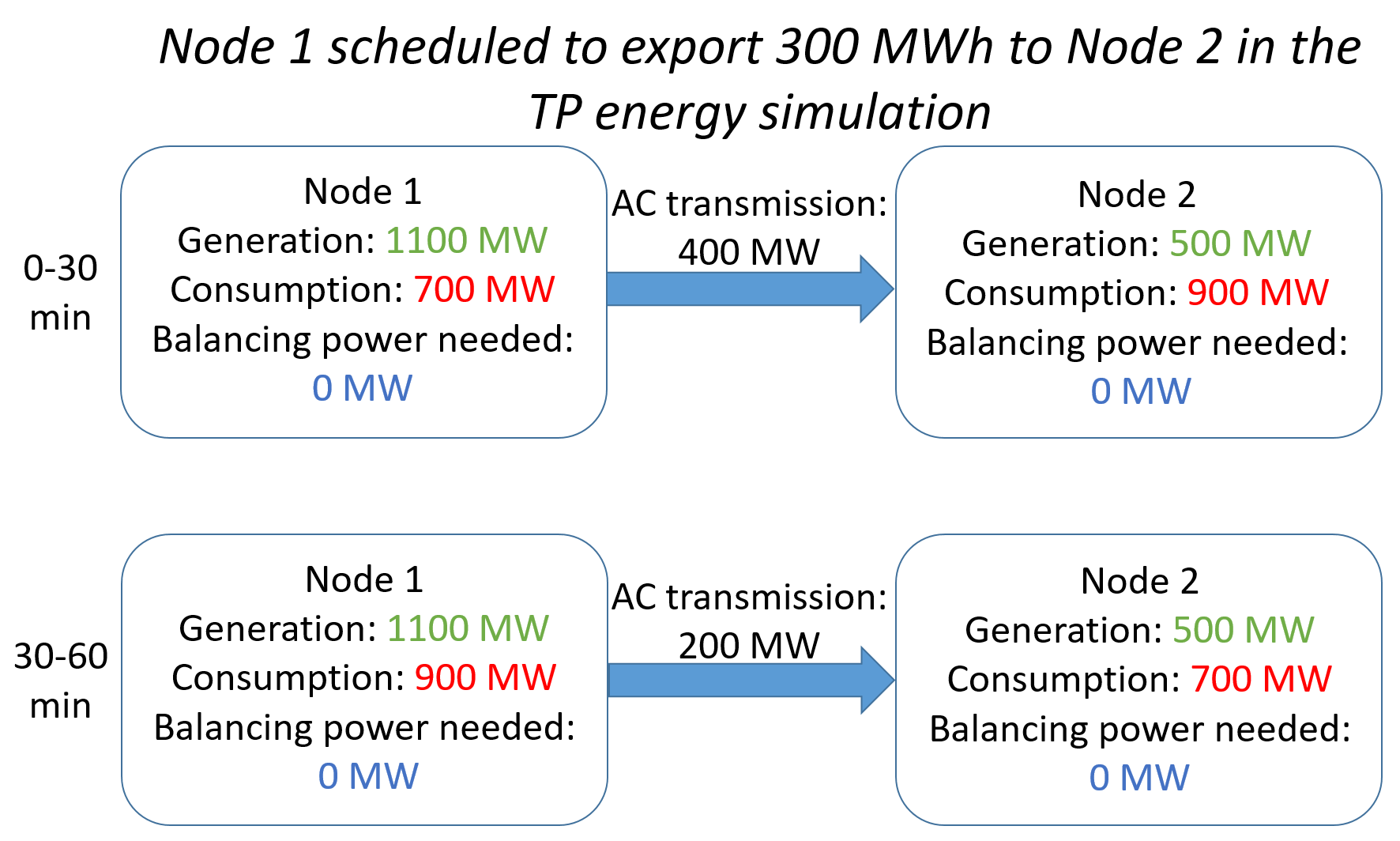}}
\caption{An example of the \textit{need of balancing power} in a simple two node system during one TP when transmission is used to \textit{net imbalances}.}
\label{fig:minimised_balancing}
\end{figure}

The \textit{netting of imbalances} is defined as the usage of AC transmission in interconnections between nodes within the studied power system to minimise the \textit{need of balancing power} in the entire power system. This allows \textit{power imbalances} causing a surplus in one node making up for \textit{power imbalances} causing a deficit in another node, given that enough transmission capacity is available between the two nodes.  However, the assumption about perfect forecasts in the \textit{TP energy simulations} also apply to AC transmission. Hence the total energy transmitted in each AC interconnection during each TP must be the same as in the \textit{TP energy simulations} even when AC transmission is used to \textit{net imbalances}. Fig.~\ref{fig:prenetting} shows the \textit{need of balancing power} in a simple two-node system with 30 minutes resolution when AC transmission is not used to \textit{net imbalances} and Fig.~\ref{fig:minimised_balancing} shows the same system when AC transmission is used to \textit{net imbalances} (assuming enough transmission capacity is available). By comparing the two figures, one can see that the \textit{netting of imbalances} eliminates the \textit{need of balancing power} in both nodes during the entire TP in this case. One should note that the \textit{netting of imbalances} will occur by nature in a synchronous power system as there always is a continuous balance between the power injected to and absorbed by the synchronous system. However, to which extent \textit{netting of imbalances} is included when planning and activating balancing services is important. An underestimation of the power system's capability to \textit{net imbalances} may lead to over-dimensioning of balancing services. An overestimation of the system's capability to \textit{net imbalances} may instead lead to transmission lines being overloaded. 
\newline

\begin{figure}[tp]
\centerline{\includegraphics[width=0.95\linewidth]{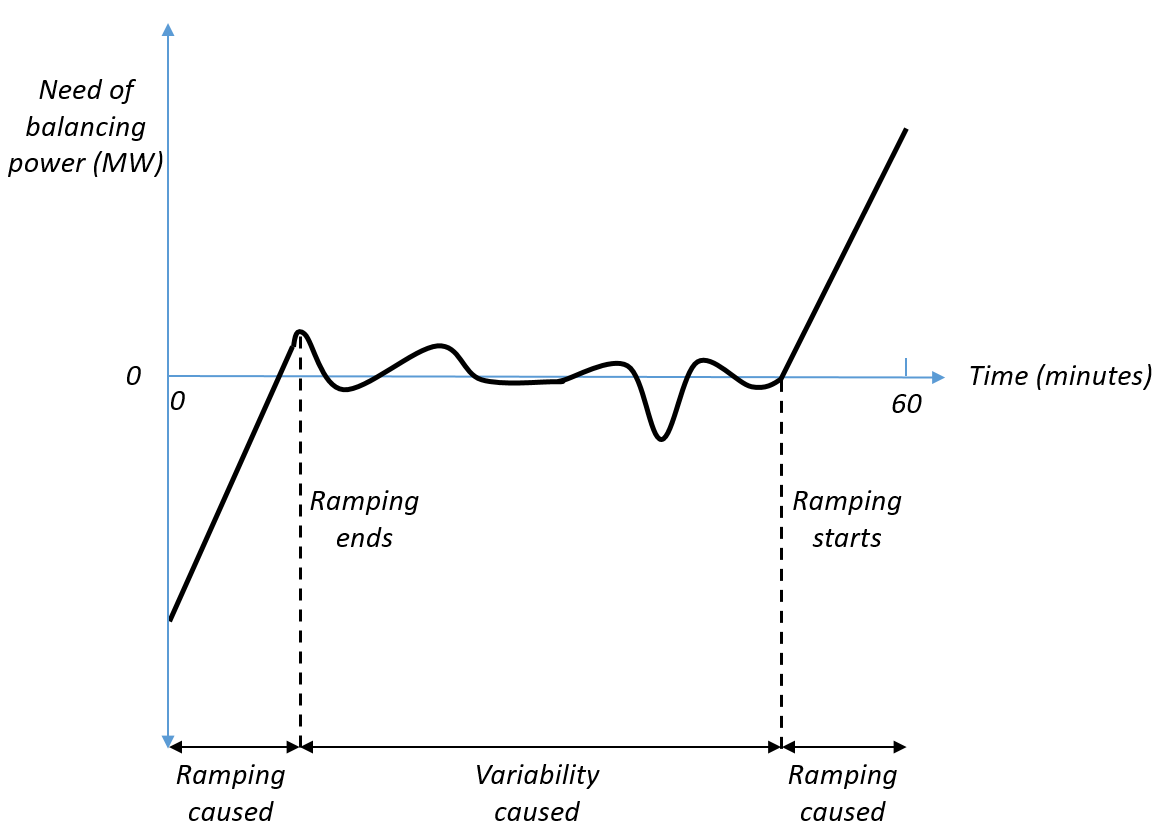}}
\caption{An example of how \textit{ramping caused needs of balancing power} are distinguished from \textit{variability caused needs of balancing power}.}
\label{fig:rampvar}
\end{figure}

Finally, we define the two main types of causes leading to a \textit{need of balancing power} if perfect TP energy forecasts are assumed. A \textit{ramping caused need of balancing power} occurs in conjunction with shifts of TP when controllable production or HVDC transmission ramps to reach its new scheduled output or interchange. If the ramping of different units do not cohere with each other, a significant \textit{need of balancing power} will temporarily occur around the shift of TP until each unit reaches its scheduled output or interchange. In contrast, a \textit{variability caused need of balancing power} may occur during the entire TP and is caused by a lack of coherence between fluctuations in non-controllable generation and demand. This most often causes smaller but more frequently occurring \textit{needs of balancing power} that is over-shadowed by the \textit{ramping caused need of balancing power} at shifts of TP. Hence, we simplify by claiming the \textit{need of balancing power} being \textit{variability caused} when there is no ramping of HVDC interconnections or controllable generation and claiming the \textit{need of balancing power} solely being \textit{ramping caused} when there clearly is ramping occurring at a shift of TP. Fig.~\ref{fig:rampvar} shows a simple example of how \textit{ramping caused needs of balancing power} are distinguished from \textit{variability caused needs of balancing power} for one TP of 60 minutes.

\section{Model description} \label{sec:model}
\begin{figure*}[tbp]
\centerline{\includegraphics[width=0.95\linewidth]{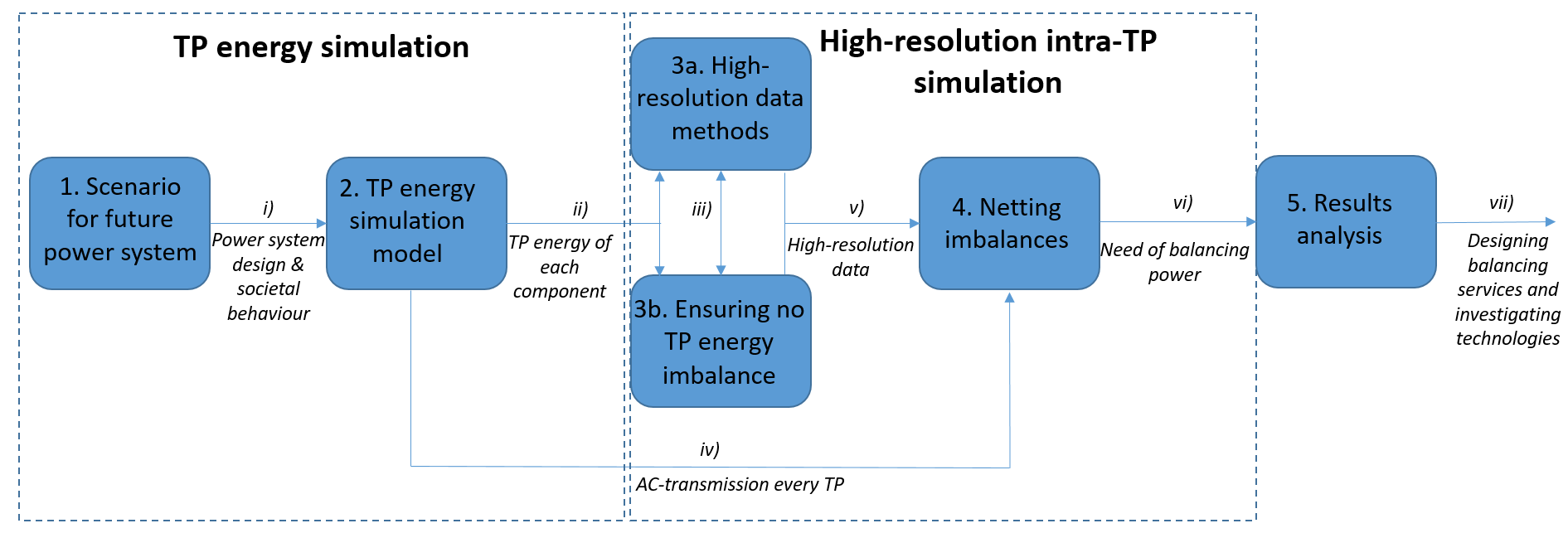}}
\caption{The process of obtaining the results in this paper from setting up a future power system scenario.}
\label{fig: workflow}
\end{figure*}

The complete process of obtaining the \textit{need of balancing power} is described in Fig.~\ref{fig: workflow}. This section presents the model used for \textit{high-resolution intra-TP simulations} that estimates the \textit{need of balancing power} by using \textit{TP energy simulation} data as input.

\subsection{Input data}
The input data to the proposed model is output data from \textit{TP energy simulations} as indicated by link ii) in Fig.~\ref{fig: workflow}. This means that the \textit{TP energy simulations} in box 2. of a power system scenario in box 1. must first be performed to use the proposed model. The input data should be given as one time series for each component in the studied system containing the energy for every simulated TP. The different types of components considered in this paper can be seen by studying the components in (\ref{eq:balpow}).

\subsection{Controllable components}
In this paper, we define controllable components as components where the power generated, consumed or transmitted is controllable in real time by an operator. This includes e.g. conventional power plants, reservoir hydro power, batteries and HVDC links. The most likely behaviour of controllable components would be aiming at providing the \textit{basic power} during each TP to avoid \textit{power imbalances} causing \textit{TP energy imbalances}. At shifts of TP, controllable components can only ramp at a limited rate and hence \textit{power imbalances} will occur around the shifts of TP. The ramping could be modelled in different ways depending on the assumptions made. The ramp rate could either be a fixed increase in power or depend on a fixed ramping period. The ramping period might also occur before, after or during the TP shift depending on factors like electricity price or financial compensation for ramping at a certain time. In this paper, we use a method named $HR_{C}$ (High-resolution controllable) to create high-resolution data of controllable components. The method is included in box 3a. in Fig.~\ref{fig: workflow}. 
\newline

Method $HR_{C}$ is used to transform data of controllable components with resolution of one TP, $t$, to data with higher resolution, $\hat{t}$. The components are modelled to provide the \textit{basic power} within each TP and to have a ramping period beginning $C_{t \rightarrow t+1}$ minutes before the shift of TP and ending $C_{t \rightarrow t+1}$ minutes after the shift of TP. During the ramp, data is fitted to a curve using linear interpolation. The parameter $C_{t \rightarrow t+1}$ is calculated for each shift of TP according to (\ref{eq:k}) if the ramp rate, $r$, is given based on how large share of the installed capacity a component can ramp during a certain time or according to (\ref{eq:m}) if the ramp rate, $r$, is given as a certain power a component changes the output with during a certain time. The parameter $C_{t \rightarrow t+1}$ depends on both input data and the assumed ramp rate. The parameter $C_{t \rightarrow t+1}$ is restricted to not be larger than half a TP (leading to a ramping period of one TP). If the parameter would be larger, there is a risk that data points created from the previous TP is overwritten or that data to be created for the following TP will overwrite the data just created. Hence, it would be impossible to ensure no \textit{TP energy imbalances} occurring. If $C_{t \rightarrow t+1}$ is set to its largest allowed value, it can be interpreted as that the ramping capability of a controllable component is over-estimated in the \textit{TP energy simulations}, thus it is of interest to analyse how often this actually happens. Method $HR_{C}$ is denoted as $HR_{C}(*)$ and (\ref{eq:HRC}) shows how a $t$-resolution time series of a component, $w$, is used to create a $\hat{t}$-resolution time series, $\hat{w}$, with method $HR_{C}$.

\begin{equation}\label{eq:k}
C_{t \rightarrow t+1} = \min \bigg(\frac{|w_{t+1} - w_{t}|}{r \cdot w^{max}} \cdot \frac{1}{2}, \frac{\hat{T}}{T} \cdot \frac{1}{2} \bigg)
\end{equation}

\begin{equation}\label{eq:m}
C_{t \rightarrow t+1} = \min \bigg( \frac{|w_{t+1} - w_{t}|}{r} \cdot \frac{1}{2}, \frac{\hat{T}}{T} \cdot \frac{1}{2} \bigg)
\end{equation}

\begin{equation}\label{eq:HRC}
\hat{w} = HR_{C}(w)
\end{equation}

\subsection{Varying components}
As opposed to controllable components, varying components depend on instantaneous weather conditions or other non-controllable factors, e.g. the decisions of individuals. Electricity consumption and vRES production are examples of varying components. If wind and solar power production are separate data categories and a smaller geographical scale in a nodal model is used, solar and wind power production may be modelled in more detail to better capture the fluctuating nature of these types of power production. A method to create high-resolution solar power production data is presented in \cite{b18} and methods to create high-resolution production data from wind farms are presented in \cite{b19} and \cite{b20}. The variability of electricity demand could also be modelled in more detail if assumptions regarding societal behaviour are made, e.g. \cite{b21} presents a model to create high-resolution domestic electricity demand data. However, when modelling larger geographical areas (like trading areas) the local high-frequency fluctuations of variable electricity production and demand most likely have a limited impact on the total area output. Hence, in this paper we use a method creating smoother time-series of varying components, named method $HR_{V}$ (High-resolution varying). The method is included in box 3a. in Fig.~\ref{fig: workflow}.
\newline

Method $HR_{V}$ is used to transform varying component data with resolution of one TP, $t$, to data with higher resolution, $\hat{t}$. This is made by letting the data follow a cubic spline curve between each step-change. The spline interpolation gives a realistic and smooth shape to the fitted curve that mimics the continuous variability of varying components better than linear interpolation. The $t$-resolution data is used in a cubic spline interpolation solution developed by \cite{b22}. As input to the interpolation solution, $t$-resolution data is used in between each step-change, i.e. $t + 0.5$. However, the spline interpolation does not ensure that no \textit{TP energy imbalances} will occur. The method $HR_{V}$ is denoted as $HR_{V}(*)$ and (\ref{eq:L}) shows how a $t$-resolution time series of a varying component, $w$, is used to create a $\hat{t}$-resolution time series, $\hat{w}$, with method $HR_{V}$.

\begin{equation}\label{eq:L}
\hat{w} = HR_{V}(w)
\end{equation}

\subsection{Ensuring no TP energy imbalances occur}
When using the data-processing methods $HR_{C}$ and $HR_{V}$ a \textit{TP energy error} may occur meaning that the energy of \textit{high-resolution intra-TP simulation} data during one TP is not equal to the energy in the \textit{TP energy simulations} for the same component. Hence, the assumption about perfect forecasts and no \textit{TP energy imbalances} in the \textit{TP energy simulations} would not be valid and thus there is a need for a method to ensure no \textit{TP energy imbalances} will occur. The method is represented by box 3b. in Fig.~\ref{fig: workflow} and as indicated by link iii), it requires the usage of method $HR_{C}$ and $HR_{V}$ iteratively. The method used was first presented in \cite{b15} and could be described as follows:
\newline

First, the difference, $h^{i}$, between the \textit{high-resolution intra-TP simulation} data, $\hat{w}^{i}$ and \textit{TP energy simulation} data, $w$, is calculated for each TP for one component by using (\ref{eq:h}) for every simulated TP in iteration $i$.

\begin{equation}\label{eq:h}
h_{t}^{i} = w_{t} - \frac{T}{\hat{T}} \sum_{\mathclap{\hat{t}=\frac{\hat{T}}{T}(t-1)+1}}^{\frac{\hat{T}}{T}t} \hat{w}_{\hat{t}}^{i}, \hspace{5pt} t = 1, \dots ,T 
\end{equation}

Second, the total error, $e^{i}$, is calculated by using (\ref{eq:error}) for a certain component in iteration $i$.

\begin{equation}\label{eq:error}
e^{i} = \frac{1}{2} \sum_{t=1}^{T} (h_{t}^{i})^{2}
\end{equation}

If the total error is not smaller than an accepted error, $e_{min}$, an updated TP-resolution data series, $a^{i+1}$, will be computed based on the TP-resolution data series in the previous iteration and the computed difference in the previous iteration according to (\ref{eq:updatelow}). The accepted error is set to an arbitrarily low number causing a negligible \textit{TP energy imbalance} in comparison to the size of the power system and the simulated time period.

\begin{equation}\label{eq:updatelow}
a^{i+1}_{t} = a^{i}_{t} + h_{t}^{i}, \hspace{5pt} t = 1, \dots , T
\end{equation}

Then, an updated high-resolution data series, $\hat{w}^{i+1}$, is created with method $HR_{C}$ or $HR_{V}$ depending on the data category by using (\ref{eq:HD}). Note that method $HR_{C}$ or $HR_{V}$ is here denoted as $f(*)$. 

\begin{equation}\label{eq:HD}
\hat{w}^{i+1} = f(a^{i+1})
\end{equation}

The method is then iterated until the total error has reached an accepted minimum level. Initially, the low-resolution time-series, $a^{1}$, is set as the time series from the \textit{TP energy simulations}, $w$. As seen in link v) in Fig.~\ref{fig: workflow}, high-resolution data not causing any \textit{TP energy imbalances} for each component is then used in box 4. where the \textit{need of balancing power} is determined after \textit{netting imbalances}. The iterative process ensuring no \textit{TP energy imbalances} occur is described in Algorithm 1.
\newline

\begin{center}
\begin{tabular}{l}
\hline
\textbf{Algorithm 1 }\\
\hline
\textbf{Data:} Low-resolution data $w$ \\
\textbf{Result:} High-resolution data $\hat{w}$\\
\textbf{Initialization:} $a^{1} = w$\\
\textbf{while} $e^{i} > e_{min}$ \textbf{do:} \\
\hspace*{5.0pt} $a^{i+1} = a^{i} + h^{i}$ \\
\hspace*{5.0pt} $\hat{w}^{i+1} = f(a^{i+1})$ \\
\hspace*{5.0pt} $h^{i+1} = w_{t} - \frac{T}{\hat{T}} \sum_{\hat{t}=\frac{\hat{T}}{T}(t-1)+1}^{\frac{\hat{T}}{T}t} \hat{w}_{\hat{t}}^{i}$ \\
\hspace*{5.0pt} $e^{i+1} = \frac{1}{2} \sum_{t=1}^{T} h_{t}^{2}$ \\
\hspace*{5.0pt} $i = i + 1$ \\
\hline
\newline
\end{tabular}
\end{center}

\begin{figure}[bp]
\centerline{\includegraphics[width=\linewidth]{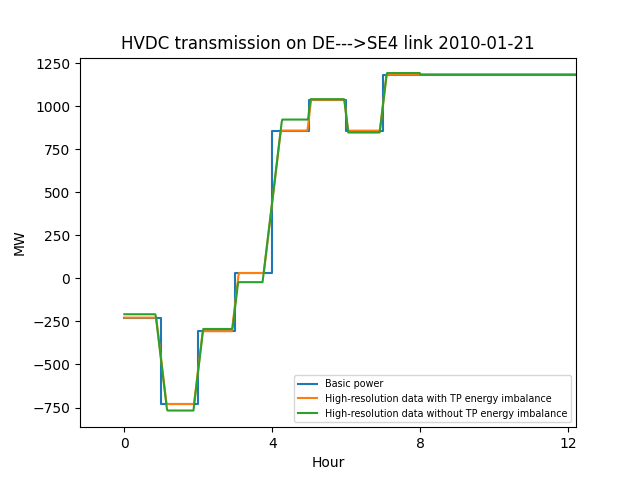}}
\caption{An example of HVDC transmission data showing the basic power and the high-resolution data processed with method $HR_{C}$ when it is ensured no \textit{TP energy imbalances} occurs and when it is not ensured any \textit{TP energy imbalances} will occur.}
\label{fig:HRC}
\end{figure}

Fig.~\ref{fig:HRC} shows an example of how HVDC transmission data changes through the different steps in the proposed method to create high-resolution data. From the \textit{TP energy simulations} in box 2. in Fig.~\ref{fig: workflow} we only have the \textit{basic power} following a step-function. By solely using method $HR_{C}$ (represented by box 3a. in Fig.~\ref{fig: workflow}) ramps will be included at TP shifts and after the ramping period the data will be the same as the \textit{basic power}. When it is then ensured no \textit{TP energy imbalances} will occur (represented by box 3b. in Fig.~\ref{fig: workflow}) the data might then deviate from the \textit{basic power} outside of the ramping period at certain occurrences. If the \textit{basic power} of an HVDC interconnection is already at its maximum transmission capacity in any direction, this then indicates that the method ensuring no \textit{TP energy imbalances} might violate transmission limits which is a drawback of this method that comes with the assumption of perfect forecasts in the \textit{TP energy simulations}.
\newline

\begin{figure}[tp]
\centerline{\includegraphics[width=\linewidth]{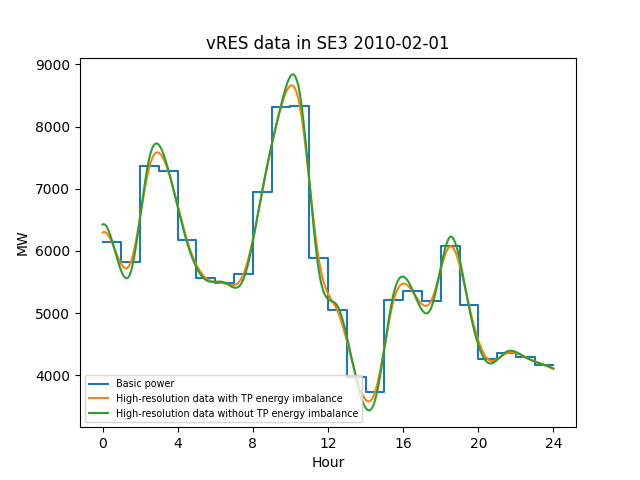}}
\caption{An example of vRES data showing the basic power and the high-resolution data processed with method $HR_{V}$ when it is ensured no \textit{TP energy imbalances} occurs and when it is not ensured any \textit{TP energy imbalances} will occur.}
\label{fig:HRV}
\end{figure}

Fig.~\ref{fig:HRV} shows an example of how vRES data changes through the different steps of the proposed method to create high-resolution data as described for Fig.~\ref{fig:HRC}. As mentioned the spline interpolation does not ensure no \textit{TP energy imbalances} occur and thus data series of varying components also changes when it is ensured no \textit{TP energy imbalances} will occur.

\subsection{Netting imbalances}
To determine the \textit{need of balancing power} when using AC transmission to \textit{net imbalances}, an optimisation problem was set up with the objective to minimise the total absolute \textit{need of balancing power} in the entire power system during the simulated period. The AC transmission between nodes is a variable that aims to \textit{net imbalances} between nodes and is modelled with a simple transport model assuming no additional transmission losses will occur compared to the \textit{TP energy simulations}. The optimisation algorithm may be computationally burdensome if using a detailed nodal model and the recommendation is then to aggregate clusters of geographically adjacent nodes to single nodes. The method is represented by box 4. in Fig.~\ref{fig: workflow} and as seen in link iv) and v), the TP energy of AC transmission corridors as well as high-resolution data for every other component in the simulated system is used as input data to the optimisation algorithm. As seen in link vi), the output data is the \textit{need of balancing power} that could then be used for further analysis of the balancing of future power systems in box 5.
\newline

The objective function to be minimised is the sum of the absolute \textit{need of balancing power} in each node during the simulated period. The objective function also includes a term where the absolute deviation of $\hat{t}$-resolution AC transmission, $|\hat{z}^{ac}_{\hat{t}+1} - \hat{z}^{ac}_{\hat{t}}|$, has a small cost, $\alpha$. The reason for this is to avoid unrealistic fluctuations of the \textit{need of balancing power} in the case of multiple interconnected nodes ending up with a surplus or deficit and the transmission lines between the nodes not being congested. If the term was not added, the total system \textit{need of balancing power} would remain the same but the \textit{need of balancing power} would be placed in a random node every time step, hence it would be harder to analyse the patterns of the \textit{need of balancing power} in each node. The objective function is described by (\ref{eq:obj}).

\begin{equation}\label{eq:obj}
\sum_{\hat{t}=1}^{\hat{T}} \sum_{n \in N} (|\hat{w}^{bal}_{\hat{t},n}|) + \alpha \sum_{\hat{t}=1}^{\hat{T}-1} \sum_{a \in A} (|\hat{z}^{ac}_{\hat{t}+1,a} - \hat{z}^{ac}_{\hat{t},a}|)
\end{equation}

In (\ref{eq:con1}), the transmission limits of the AC transmission are set as $\overline{Z}^{ac}$ and $\underline{Z}^{ac}$ which represents the  largest allowed flow in each direction of an AC interconnection.

\begin{equation}\label{eq:con1}
\underline{Z}^{ac}_{a} \leq \hat{z}_{\hat{t},a} \leq \overline{Z}^{ac}_{a}, \hspace{5pt}\hat{t} = 1, \dots, \hat{T}, \hspace{5pt} a \in A
\end{equation}

In (\ref{eq:con2}), the energy of $\hat{t}$-resolution AC transmission every TP is constrained to be equal to the energy in the \textit{TP energy simulation} to ensure no \textit{TP energy imbalances} will occur. This constraint is explained by the assumption of perfect forecasts in the \textit{TP energy simulations} and that no \textit{TP energy imbalances} will occur. 

\begin{equation}\label{eq:con2}
\frac{T}{\hat{T}} \sum_{\mathclap{\hat{t}=\frac{\hat{T}}{T}(t-1)+1}}^{\frac{\hat{T}}{T}t} \hat{z}^{ac}_{\hat{t},a} = z^{ac}_{t,a}, \hspace{5pt} t = 1, \dots ,T, \hspace{5pt} a \in A
\end{equation}

The objective function is also subject to a balance constraint which requires power balance in every node at every time instance during the simulated period meaning that the power deficit must equal the \textit{need of balancing power} in each node. Here the high-resolution data-series of different components created with methods $HR_{C}$ and $HR_{V}$ are added together for each node together with import and export from AC transmission to get the power deficit which should equal the \textit{need of balancing power}. The balance constraint is described in (\ref{eq:con3}).

\begin{align}\label{eq:con3}
&-\hat{g}^{hy}_{\hat{t},n} - \hat{g}^{fl}_{\hat{t},n} - \hat{g}^{th}_{\hat{t},n} - \hat{g}^{nu}_{\hat{t},n} - \hat{g}^{res}_{\hat{t},n} + \hat{d}_{\hat{t},n} + \sum_{\mathclap{a \in \{A:n1=n\}}}\hat{z}^{ac}_{\hat{t},a} \\
&  \text{\hspace{2pt}} - \text{\hspace{2pt}} \sum_{\mathclap{a \in \{A:n2=n\}}}\hat{z}^{ac}_{\hat{t},a} \text{\hspace{2pt}} + \text{\hspace{2pt}} \sum_{\mathclap{b \in \{B:n1=n\}}}\hat{z}^{dc}_{\hat{t},b} \text{\hspace{2pt}} - \text{\hspace{2pt}} \sum_{\mathclap{b \in \{B:n2=n\}}}\hat{z}^{dc}_{\hat{t},b} = \hat{w}_{\hat{t},n}^{bal}, \nonumber \\
&\hat{t} = 1, \dots ,\hat{T}, \hspace{5pt} n \in N \nonumber 
\end{align}

The linear optimisation problem determining the \textit{need of balancing power} in each node and the high-resolution AC transmission between nodes is described as:
\begin{align*}
\centering
    \text{minimise (\ref{eq:obj}), }&\text{subject to:} \\
    \text{(\ref{eq:con1}) }&\text{transmission limits} \\
    \text{(\ref{eq:con2}) }&\text{hourly transmission} \\
    \text{(\ref{eq:con3}) }&\text{power balance} \\
\end{align*}

\section{Case study} \label{sec:case}
The input data to the model for the case study are simulation results from the \textit{TP energy simulations} in \cite{b5} of the scenario "Electrification renewables year 2045". In this scenario, the Nordic power system has nearly 100\% renewable electricity generation and an almost twice as high electricity demand compared to today (2022) with an electricity consumption of 707 TWh/year as e.g. hydrogen use and electric vehicles have become important parts of the Nordic society. The \textit{TP energy simulations} in \cite{b5} is a cost minimisation problem using the models BID3 and EMPS to find the cost-optimal way to meet the electricity demand of the entire North European power system every single hour during each simulated year. Historical weather data for 35 years and stochastically occurring outages in components together with a number of assumptions regarding how the power system will be designed and operated in future scenarios are used as input to the model. Except from the "Electrification renewables" scenario, three other scenarios are simulated in \cite{b5} for both years 2035 and 2045. However, none of these scenarios are investigated in this paper as the sensitivity analysis focuses on assumptions in the \textit{high-resolution intra-TP simulations} rather than assumptions in the \textit{TP energy simulations}. In this paper, we use simulation results with weather data from February and July in 2010 to analyse the \textit{need of balancing power} during a cold winter month when the electricity demand in general is high and a warm summer month when the electricity demand in general is low. The input data to the model are given as time series with the TP energy for each component in the unit of \textit{MWh/h} for each of the eleven trading areas in the Nordic synchronous power system. Thus, $t$ is always set as 60 minutes. Also, ramp rates of controllable components must be provided to the model as well as transmission capacities of AC interconnections. The data series provided from the \textit{TP energy simulations} in \cite{b5} are categorised as follows: 
\newline

\begin{itemize}
    \item Controllable components
    \begin{itemize}
        \item HVDC transmission
        \item Thermal production
        \item Hydro production
        \item Nuclear production
        \item Short-term flexibility (demand side response, batteries, etc.)
    \end{itemize}
    \item{Varying components}
    \begin{itemize}
        \item Demand
        \item vRES production
    \end{itemize}
    \item Other
    \begin{itemize}
        \item AC transmission
        \newline
    \end{itemize}
\end{itemize}

When determining the \textit{need of balancing power} by performing the \textit{high-resolution intra-TP simulations}, $\hat{t}$ is always set as one minute in this case study as minutely time series of the \textit{need of balancing power} are granular enough to use for further studies while still being produced at a computationally low burden. However, the value of the $\hat{t}$-parameter could be set to an optional value that suits the aim of one's studies. Solving the model of the Nordic synchronous power system for one month with hourly data as input and minutely data as output with Gurobi 9.1. in Python on a laptop  with an 11th Gen Intel(R) Core i7-1185G7 @ 3.00GHz processor and 32 GB of RAM takes about 300 to 900 seconds depending on which assumptions are made and how many days there are in the studied month. 
\newline

In this case study, we investigate how the assumed ramp rate of controllable production components and the assumed transmission capacity of AC interconnections impact the \textit{need of balancing power} by varying these parameters. Regarding ramp rates of controllable production components we study two different cases, the "normal ramping case" and the "fast ramping case". The "normal ramping case" aim to reflect ramp rates used today while the "fast ramping case" aim to reflect the technically fastest possible ramping. The ramp rates for different production types in the "normal ramping case" and the "fast ramping case" are shown in Table~\ref{tab:ramps}. The ramp rates are given as how many percent of the maximum production level the component can increase or decrease its production each minute. As the maximum production level, $g_{max}$, is not given in the input data, it is set as the highest level of production during the studied period. To avoid division by zero if a unit is not producing at all during the studied period, $g_{max}$ is set to a minimum level of 1 MW/min. Short-term flexibility is assumed to have the same ramp rates as hydro power production as it may include a variety of different technologies and thus is hard to give a certain ramp rate. The "normal ramping case" ramp rate of hydro power is set as 5\%/min which is lower than the fastest ramp rate technically possible in \cite{b23}, as hydro power most often ramp slower than technically possible to avoid large \textit{power imbalances}. Ramp rates for thermal power and nuclear power were found in \cite{b24} and \cite{b25}. The ramp rate of all HVDC interconnections is always set as 30 MW/min according to existing regulations in the Nordic power system \cite{b26}. Regarding AC transmission, two different cases are investigated in this paper, either if the transmission limits, $\underline{Z}^{ac}$ and $\overline{Z}^{ac}$, are set as the net transfer capacity (NTC) in each direction or if the transmission reliability margin (TRM) is allowed to be used to \textit{net imbalances}, hence the positive/negative transmission limits are set as the positive/negative NTC plus/minus the TRM. The TRM is a certain margin that is kept between the actual transmission capacity of a transmission corridor and the NTC, which is the transmission capacity reported to the market. Thus, the TRM ensures that a short-term fluctuation from planned transmission will not cause transmission lines being overloaded. In reality, the TRM will automatically be used to \textit{net imbalances}, but assuming that the TRM is used to \textit{net imbalances} when planning and activating balancing services increases the risk of overloading transmission lines in case of additional disturbances. The TRM of each AC interconnection is assumed to remain the same as in \cite{b27}.
\newline

\begin{table}[tp]
\caption{The different ramp rates of controllable production components.}
\begin{center}
\begin{tabular}{|l|l|l|l|} \hline
  & Hydro, Flexibility & Thermal & Nuclear \\
  & [\%$g_{max}$/min] & [\%$g_{max}$/min] & [\%$g_{max}$/min] \\
\hline
Normal & 5 & 3 & 1.5 \\
\hline
Fast & 15 & 10 & 5 \\
\hline
\end{tabular}
\newline
\label{tab:ramps}
\end{center}
\end{table}

In total, the two ramping cases and the two AC transmission cases lead to four different setups of assumptions ($S_{1}$-$S_{4}$) that are studied in this paper. Each setup of assumptions is described in Tab.~\ref{tab:scenarios}. As both a February month and a July month is studied, eight simulations were performed in total. The simulations ran generated minutely data series for each component including the \textit{need of balancing power} in each trading area of the Nordic synchronous system. However, we limit the analysis of simulation results in this paper to solely study the \textit{need of balancing power} in SE4 (southernmost Sweden), an area with three HVDC interconnections and almost 95\% of the total installed capacity being vRES production in the studied scenario.

\begin{table}[bp]
\caption{The different setups of assumptions that are analysed.}
\begin{center}
\begin{tabular}{|c|c|c|} \hline
  & NTC & NTC + TRM \\
\hline
Normal ramping & $S_{1}$ & $S_{2}$ \\
\hline
Fast ramping & $S_{3}$ & $S_{4}$ \\
\hline
\end{tabular}
\label{tab:scenarios}
\end{center}
\end{table}

\section{Results} \label{sec:results}
\begin{figure}[tbp]
\centerline{\includegraphics[width=\linewidth]{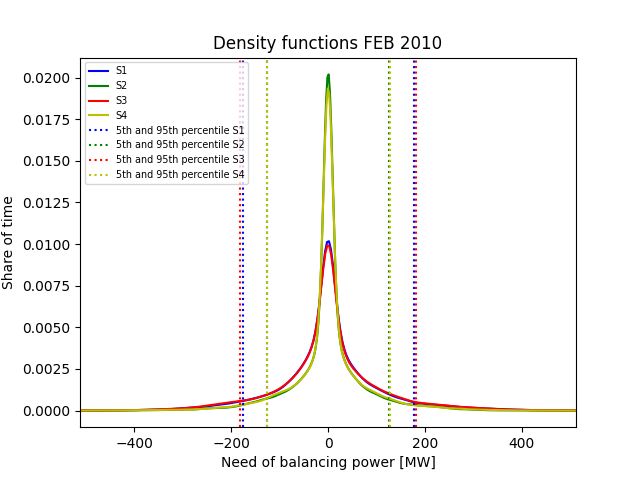}}
\caption{Density plot of the need of balancing power during the simulated February month for setups $S_{1}$-$S_{4}$ in SE4.}
\label{fig:febdensity}
\end{figure}

\begin{figure}[tbp]
\centerline{\includegraphics[width=\linewidth]{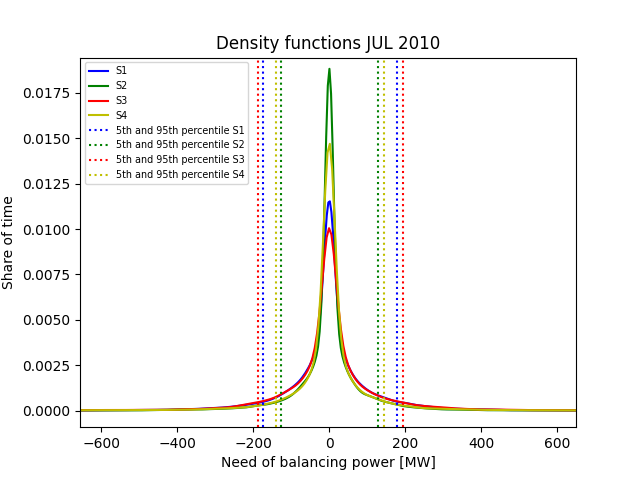}}
\caption{Density plot of the need of balancing power during the simulated July month for setups $S_{1}$-$S_{4}$ in SE4}
\label{fig:juldensity}
\end{figure}

Key results regarding the simulations of setups $S_{1}$-$S_{4}$ are presented in Tab.~\ref{tab:febresults} for the studied February month and Tab.~\ref{tab:julresults} for the studied July month. The density plots of the \textit{need of balancing power} for all setups are shown in Fig.~\ref{fig:febdensity} for the February month and in Fig.~\ref{fig:juldensity} for the July month. The results show there will be a significant \textit{need of balancing power} with peaks exceeding 1000 MW of both positive and negative \textit{needs of balancing power} in many of the studied setups of assumptions in both the studied months. In the studied February month, the peaks are lower for setups $S_{2}$ and $S_{4}$ compared to setups $S_{1}$ and $S_{3}$. In the studied July month, the peaks are instead lower for setups $S_{1}$ and $S_{2}$ compared to setups $S_{3}$ and $S_{4}$. The mean absolute \textit {need of balancing power} varies between 42 and 68 MW in the different setups of assumptions in the studied February month and between 44 and 73 MW in the different setups of assumptions in the studied July month. In both months, the mean absolute \textit{need of balancing power} is lower for setups $S_{2}$ and $S_{4}$ compared to setups $S_{1}$ and $S_{3}$. There is no \textit{need of balancing power} for just over 20\% of the time in setups $S_{2}$ and $S_{4}$ in both the studied February and July month while the share of time with no \textit{need of balancing power} for setups $S_{1}$ and $S_{3}$ is about 11\% for the studied February month and about 13\% for the studied July month. From the density plots it can be seen that the setups $S_{2}$ and $S_{4}$ significantly reduces the \textit{need of balancing power} compared to setups $S_{1}$ and $S_{3}$ during both studied months. However, during the studied July month there is a clearer difference between setups $S_{2}$ and $S_{4}$ as well as between $S_{1}$ and $S_{3}$ compared to the studied February months. In July, $S_{2}$ leads to a reduced \textit{need of balancing power} compared to $S_{4}$ and the \textit{need of balancing power} in $S_{1}$ is reduced compared to $S_{3}$.  
\newline

\begin{table}[tp]
\caption{Key results regarding the need of balancing power in SE4 during the simulated February month.}
\centering
\begin{tabular}{|c|c|c|c|c|} \hline
 & max $\hat{w}^{bal}_{SE4}$ & min $\hat{w}^{bal}_{SE4}$ & $\mu(|\hat{w}^{bal}_{SE4}|)$ & $\hat{w}^{bal}_{SE4}=0$ \\
 & [MW] & [MW] & [MW] & [\% of time] \\ 
\hline
$S_{1}$ & 1 093 & - 902 & 67 & 11.28 \\
\hline
$S_{2}$ & 486 & - 842 & 42 & 20.81 \\
\hline
$S_{3}$ & 1 102 & - 923 & 68 & 10.73 \\
\hline
$S_{4}$ & 517 & - 910 & 44 & 20.74 \\
\hline
\end{tabular}
\label{tab:febresults}
\end{table}

\begin{table}[tp]
\caption{Key results regarding the need of balancing power in SE4 during the simulated July month.}
\centering
\begin{tabular}{|c|c|c|c|c|} \hline
 & max $\hat{w}^{bal}_{SE4}$ & min $\hat{w}^{bal}_{SE4}$ & $\mu(|\hat{w}^{bal}_{SE4}|)$ & $\hat{w}^{bal}_{SE4}=0$ \\
 & [MW] & [MW] & [MW] & [\% of time] \\ 
\hline
$S_{1}$ & 1 046 & - 974 & 64 & 13.13 \\
\hline
$S_{2}$ & 926 & - 1 211 & 44 & 20.23 \\
\hline
$S_{3}$ & 1 575 & - 1 558 & 73 & 12.74 \\
\hline
$S_{4}$ & 1 522 & - 1 500 & 53 & 20.31 \\
\hline
\end{tabular}
\label{tab:julresults}
\end{table}

\begin{figure}[htbp]
\centerline{\includegraphics[width=\linewidth]{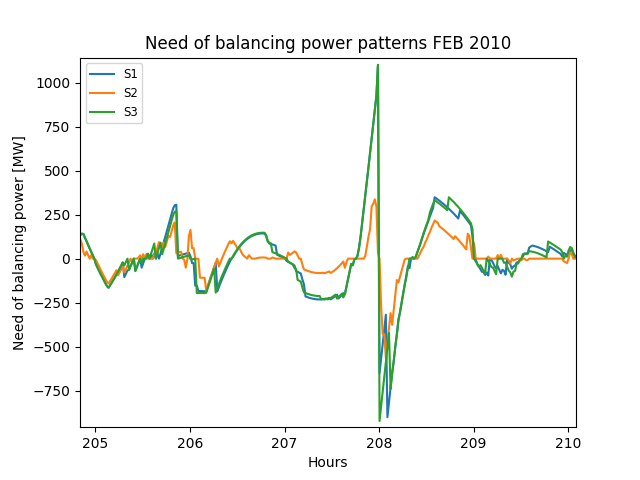}}
\caption{The \textit{need of balancing power} during 5 hours in the simulated February month for scenario $S_{1}$-$S_{3}$ in SE4.}
\label{fig:febpatterns}
\end{figure}

In Fig.~\ref{fig:febpatterns} the minutely \textit{need of balancing power} is shown for 5 hours during the simulated February month for setups $S_{1}$-$S_{3}$. It can be seen that the peaks of $S_{2}$ are smaller than the peaks of the other setups in the plot. Overall, the plots of $S_{1}$ and $S_{3}$ are very similar to each other. In all three setups, there is a pattern of the largest peaks of a \textit{need of balancing power} being \textit{ramping caused} occurring around TP shifts, e.g. at the shift from hour 207 to hour 208. The \textit{variability caused need of balancing power} within TPs is at certain occasions high as well, e.g. in the middle of hour 208.
\newline

In Tab.~\ref{tab:DCproblems}, the share of TP shifts where today's ramp rate of HVDC interconnections to SE4 were not adequate to ensure no \textit{TP energy imbalances} would occur is presented, i.e. the model had to set a higher ramp rate to ensure no \textit{TP energy imbalance} did occur. It can be seen that this occurs frequently for especially the interconnection between Germany and SE4 (DE $\rightarrow$ SE4) with the ramp rate not being adequate almost 3\% of the TP shifts during the studied July month.

\begin{table}[bp]
\caption{Showing how often larger ramp rates than today's were required for the HVDC links connected to SE4.}
\centering
\begin{tabular}{|c|c|c|c|} \hline
 & DE $\rightarrow$ SE4 & LT $\rightarrow$ SE4 & PL $\rightarrow$ SE4 \\
 & [\% of TP shifts] & [\% of TP shifts] & [\% of TP shifts] \\ 
\hline
February & 1.79 & 0.45 & 0.30 \\
\hline
July & 2.96 & 0.00 & 0.54 \\
\hline
\end{tabular}
\label{tab:DCproblems}
\end{table}

\section{Discussion} \label{sec:discussion}
The results showed a \textit{need of balancing power} will occur most of the time in all studied setups during both the studied months in SE4, even though perfect forecasts for the energy during each TP were assumed. The peak \textit{needs of balancing power} were significantly high but during most of the time, the \textit{need of balancing power} was not near the magnitude of the peaks. In general, the largest \textit{needs of balancing power} did occur at TP shifts being \textit{ramping caused} while the \textit{variability caused needs of balancing power} tended to be smaller but still continuously occurring and not of a negligible size. Overall, the \textit{need of balancing power} surprisingly did tend to be larger during the studied July month than the studied February month. This may be explained by solar power production being higher during summer causing larger fluctuations in vRES production and thus other components need to ramp more heavily to avoid \textit{TP energy imbalances}.
\newline

It could be seen that allowance of using the TRM to \textit{net imbalances} helps shaving peaks of the \textit{need of balancing power} and significantly reduced the mean \textit{need of balancing power} as well as how large share of the time there was a \textit{need of balancing power} for both studied months by comparing the results of $S_{2}$ and $S_{4}$ with $S_{1}$ and $S_{3}$. In reality the TRM will be used to \textit{net imbalances} by nature, but one can question if this should be relied upon when dimensioning balancing services as the risk of overloading transmission lines in occurrence of disturbances increases if the TRM is not available to handle such disturbances. The assumption regarding fast ramping of controllable components in setups $S_{3}$ and $S_{4}$ had a larger impact in the studied July month than the studied February month where especially the peak positive and negative \textit{needs of balancing power} were clearly increased by the assumption of fast ramping in the July month. As mentioned, this may be explained by higher vRES production during summer causing larger fluctuations forcing other components to ramp more heavily between TPs, if this is not made in a coordinated manner the \textit{power imbalances} of different component may not cohere causing large \textit{needs of balancing power}.
\newline

Some assumptions had to be made to be able to run the model based on the data available, namely about the ramp rates of controllable components. As shown, the assumed ramp rate of HVDC interconnections were not always adequate to ensure no \textit{TP energy imbalances} occurring. This can either be interpreted as an over-estimation of the ramping capability in the \textit{TP energy simulations}, i.e. that the TP energy differs too much between two adjacent TPs, or that grid codes regarding ramp rates of HVDC interconnections is expected to be updated until year 2045. However, in 2045 Sweden is expected to have shifted to 15 minutes TP since long time ago. Thus, HVDC links will be allowed to ramp four times per hour instead of one time per hour and large shifts in transmitted energy per hour may be less of an issue.

\section{Conclusions} \label{sec:conclusions}
In this paper we have proposed a model to perform \textit{high-resolution intra-TP simulations} to estimate the future \textit{need of balancing power} based on \textit{TP energy simulations}. By knowing the \textit{need of balancing power}, the balancing potential of different technologies can be evaluated, balancing services can be dimensioned and technical requirements of balancing services can be designed. The model was applied to a scenario of year 2045 in the Nordic power system with nearly 100\% renewable generation and a significantly increased electricity demand. The impact of assumptions regarding AC transmission limits and ramping of controllable components on the \textit{need of balancing power} in the trading area SE4 was analysed.
\newline

The results showed positive/negative peaks of a \textit{need of balancing power} exceeding 1000 MW did occur in SE4 in both a simulated February month and a simulated July month. The peaks were most often \textit{ramping caused needs of balancing power}. Most of the time, there was a \textit{need of balancing power}, either \textit{variability caused} or \textit{ramping caused}, but not of the same magnitude as the more rarely occurring peaks. If the TRM was used to \textit{net imbalances}, the mean \textit{need of balancing power} as well as how large share of the time there was a \textit{need of balancing power} decreased. The TRM also helped in shaving peaks of \textit{ramping caused needs of balancing power}. Faster ramping of controllable components generally increased the \textit{ramping caused need of balancing power}. The assumed ramp rate of controllable components had a larger impact during the studied July month than during the studied February month.

\section{Future work} \label{sec:future}
The main area of future work is to include uncertain forecasts of demand and vRES production in the model to capture stochastic power system behaviours as these will cause a considerable \textit{need of balancing power} in the future. As of right now, the model estimates the \textit{need of balancing power} based on deterministic \textit{TP energy simulations}. Future work includes using the \textit{need of balancing power} to estimate the balancing capability of technologies like electric vehicles, hydro power, gas turbines and energy storage as well as to use the \textit{need of balancing power} to dimension future balancing services and design technical requirements for future balancing services, like frequency restoration reserves. Future work also includes investigating how the planned shift to 15 minutes TP in the Nordic power system will impact the \textit{need of balancing power}. Lastly, the modelling of controllable components in \textit{TP energy simulations} is also a topic for future work as e.g. transmission limits and assumed ramp rates of HVDC interconnections sometimes were exceeded when the scenario of the Nordic power system in 2045 was studied in an intra-TP time resolution.

\end{document}